\documentclass[smallextended]{svjour3}       
\smartqed  
\usepackage{graphicx}
\usepackage[cmex10]{amsmath}
\usepackage{amssymb}
\usepackage{amsbsy}
\usepackage{algorithmicx,algorithm}
\usepackage{algpseudocode}
\usepackage{threeparttable}
\usepackage{enumerate}
 \usepackage{mathptmx}      
\newtheorem{thm1}{Theorem} 
\newtheorem{thm2}[thm1]{Theorem}
\newtheorem{thm3}[thm1]{Theorem}

\newtheorem{lemma1}{Lemma} 
\newtheorem{lemma2}[lemma1]{Lemma}
\newtheorem{lemma3}[lemma1]{Lemma}
\newtheorem{lemma4}[lemma1]{Lemma}
\newtheorem{lemma5}[lemma1]{Lemma}
\newtheorem{lemma6}[lemma1]{Lemma}
\newtheorem{lemma7}[lemma1]{Lemma}
\newtheorem{lemma8}[lemma1]{Lemma}
\newtheorem{lemma9}[lemma1]{Lemma}
\newtheorem{lemma10}[lemma1]{Lemma}
\newtheorem{lemma11}[lemma1]{Lemma}

\newtheorem{expn1}{Example}
\newtheorem{expn2}[expn1]{Example}
\newtheorem{expn3}[expn1]{Example}

\begin{document}

\title{Two classes of linear codes with a few weights based on twisted Kloosterman sums
\thanks{Communicated by Pascale Charpin, Alexander Pott}
}


\author{Minglong Qi \and Shengwu Xiong  }


\institute{Minglong Qi   \and Shengwu Xiong 
               \at {School of Computer Science and Technology,  Wuhan University of Technology\\
                                          Mafangshan West Campus, 430070 Wuhan City, China\\}
              \email{mlqiecully@163.com (Minglong Qi)}\\
              \email{xiongsw@whut.edu.cn (Shengwu Xiong)}             
}

\date{Received: date / Accepted: date}

\maketitle

\begin{abstract}
Linear codes with a few weights have wide applications in information security, data storage systems, consuming electronics and communication systems. Construction of the linear codes with a few weights and determination of their parameters are an important research topic in coding theory. In this paper, we construct two  classes of linear codes with a few weights and determine their complete weight enumerators based on  twisted Kloosterman sums. 

\end{abstract}

\section{Introduction}

Let $ q=p^{e} $ where $ p $ is an odd prime and $ e $ is a positive integer. $ \mathbb{F}_{q} $ denotes the finite field with $ q $ elements and $ \mathbb{F}_{q}^{*}=\mathbb{F}_{q}\setminus \lbrace 0 \rbrace $. Let $ F(X)=aX^{p^{\alpha}+1} \in \mathbb{F}_{q}[X]$ be a mapping from $ \mathbb{F}_{q} $ onto itself, where $ a\in \mathbb{F}_{q}^{*} $ and $ \alpha $ is a positive integer. Let $ d=\gcd(e,\alpha)$ and $ t $ be a proper divisor of $ d $, i.e., $ 1<t<d, t\mid d $. Throughout this paper, $ e/d $ is assumed to be odd, and the meaning of $ p, e, \alpha, d $ and $ t $ are kept fixed. 

Let $ k\geq 1 $ be an arbitrary integer, and $ \chi^{(k)} $ denote the canonical additive character of $ \mathbb{F}_{p^{k}} $. In this paper, we consider the value distribution of the following exponential sum and its application in the construction of some linear codes with a few weights:
\begin{equation}\label{BSum_def}
 L_{\alpha,b}(a,u)=\sum_{z\in \mathbb{F}_{p}^{*}}\chi^{(1)}(-uz)\sum_{y\in \mathbb{F}_{p^{t}}^{*}}\chi^{(t)}(-ay)\sum_{x\in \mathbb{F}_{q}}\chi^{(e)}(yx^{p^{\alpha}+1}+zbx),
\end{equation}
where  $a\in  \mathbb{F}_{p^{t}}, u\in \mathbb{F}_{p} $ and $ b\in \mathbb{F}_{q}^{*} $.

We shall see that the explicit evaluation of (\ref{BSum_def}) involves ordinary Kloosterman sums and twisted Kloosterman sums. Let $ a,b\in \mathbb{F}_{p^{t}}^{*} $. The twisted Kloosterman sums are defined by
\begin{equation}\label{TwistedKSum_def}
T(\eta^{(t)};a,b)=\sum_{\substack{x\in\mathbb{F}_{p^{t}}^{*}}}\eta^{(t)}(x)\chi^{(t)}(bx+ax^{-1}),
\end{equation}
and the ordinary Kloosterman sums are defined by
\begin{equation}\label{KSum_def}
K(\eta^{(t)};a,b)=\sum_{\substack{x\in\mathbb{F}_{p^{t}}^{*}}}\chi^{(t)}(bx+ax^{-1}),
\end{equation} 
where $ \eta^{(t)} $ denotes the quadratic character of $ \mathbb{F}_{p^{t}}^{*} $ (see Section 3 for the details on group characters). Note that (\ref{TwistedKSum_def}) is just a variation of twisted Kloosterman sums, for more general treatment, the reader is referred to \cite{BB27,BB01,BB22}. For the divisibility of Kloosterman sums modulo an integer and their basic properties, see \cite{BB06,BB19} and the references therein. It is known that in general, the explicit evaluation of (\ref{KSum_def}) is a hard  problem. However, (\ref{TwistedKSum_def}) can be explicitly evaluated. Sometimes, we use the alias  Sali\'{e} sums to refer to (\ref{TwistedKSum_def}).

An $ [n,k,d] $ linear code $ \mathcal{C} $ over $ \mathbb{F}_{p} $ is a $ k- $dimensional subspace of $ \mathbb{F}_{p}^{n} $ with minimum (Hamming) distance $ d $. Let $ A_{i} $ denote the number of codewords with Hamming weight $ i $ of $ \mathcal{C} $. Then, the Hamming weight enumerator of $ \mathcal{C} $ is defined by
\begin{equation*}
1+A_{1}z+A_{2}z^{2}+\ldots+A_{n}z^{n}.
\end{equation*}
The sequence $ (1,A_{1},A_{2},\cdots,A_{n}) $ is called the Hamming weight distribution of the code $ \mathcal{C} $. Linear codes with a few weights have wide applications in secret sharing \cite{BB05,BB17}, authentication codes \cite{BB11}, association schemes \cite{BB04} and strongly regular graphs \cite{BB03}.

Let $ \mathbf{c}=(c_{0},c_{1},\cdots,c_{n-1}) $ be a codeword of a linear code $ \mathcal{C} $. Define the complete weight enumerator of $ \mathbf{c} $ by 
\begin{equation*}
w[\mathbf{c}]=w_{0}^{t_{0}}w_{1}^{t_{1}}\ldots w_{p-1}^{t_{p-1}},
\end{equation*}
where $ t_{i} $ is the number of coordinates of $ \mathbf{c} $ that are equal to $ c_{i} $. It is clear that $ \sum_{i=0}^{p-1}t_{i}=n $, the length of the code $ \mathcal{C} $. Then, the complete weight enumerator of the code $ \mathcal{C} $ can be defined by
\begin{equation*}
CWE(\mathcal{C})=\sum_{\substack{\mathbf{c}\in \mathcal{C}}}w[\mathbf{c}].
\end{equation*}

The complete weight enumerator of a linear code is an important parameter, and in general difficult to be evaluated. Once it is determined, the Hamming weight distribution directly follows. Blake and Kith \cite{BB02,BB21x}, in studying Reed-Solomon codes, showed that the complete weight enumerator may be useful in soft decision decoding. Helleseth and Kholosha \cite{BB19} found that the complete weight enumerator of a linear code is related to monomial and quadratic bent functions. In \cite{BB11,BB12}, Ding \textit{et al} demonstrated that the deception probabilities of some authentication codes could be calculated by applying the complete weight enumerators of these codes.

In \cite{BB07,BB13,BB14}, the authors investigated the complete weight enumerators of certain constant composition codes. In \cite{BB23,BB24}, Kuzmin and Nechaev studied the generalized Kerdok code and related linear codes over Galois ring, and determined their complete weight enumerators. Li \textit{et al} \cite{BB26} investigated the complete weight enumerator of some new linear codes using Galois theory.  For the recent development on the complete weight enumerator of linear codes, the reader is referred to \cite{BB26,BB28,BB35,BB36,BB39,BB41} and the references therein.

 Let $ D=\lbrace d_{1},d_{2},\cdots,d_{n}\rbrace $ be an $ n- $subset of $ \mathbb{F}_{q} $. Then, a linear code can be defined from the set $ D $:
\begin{equation*}
\mathcal{C}_{D}=\biggl\lbrace(\mathrm{Tr}(d_{1}b),\mathrm{Tr}(d_{2}b),\cdots,\mathrm{Tr}(d_{n}b)):b\in \mathbb{F}_{q}\biggr\rbrace,
\end{equation*}
where $ \mathrm{Tr}(x) $ denotes the absolute trace function from $ \mathbb{F}_{q} $ to $ \mathbb{F}_{p} $.
The set $ D $ is called the \textit{defining set} of the linear code $ \mathcal{C}_{D} $ (see \cite{BB17,BB18} for details). The defining set method is a generic one in the construction of linear codes. A proper definition of \textit{the definig set} may lead to linear codes optimal and with a few weights, which is a very attracting feature of this method.  Since the introduction of the defining set method, a lot of work have been devoted to construct linear codes with a few weights,
for instance, see \cite{BB33,BB35,BB39,BB41,BB42} and the references therein.

Let $ \mathrm{Tr}^{e}_{t}(x) $ denote the trace function from $ \mathbb{F}_{p^{e}} $ to $ \mathbb{F}_{p^{t}} $ where $ t\mid e $, i.e., $ \mathrm{Tr}^{e}_{t}(x)=\sum_{i=0}^{e/t-1} x^{p^{ti}}$.  Let $ a\in \mathbb{F}_{p^{t}} $. Below is the defining set used in  the present paper:
\begin{equation}\label{definig_set}
D_{a}=\biggl\lbrace x\in \mathbb{F}_{q}^{*} : \mathrm{Tr}^{e}_{t}\bigl(x^{p^{\alpha}+1}\bigr)=a \biggr\rbrace.
\end{equation}
Let the elements of $ D_{a} $ be enumerated as $ D_{a}=\lbrace d_{1},d_{2},\cdots,d_{n}\rbrace $. In this paper, we will consider two classes of linear codes  defined by:
\begin{equation}\label{linear_code_cda}
\mathcal{C}_{D_{a}}=\biggl\lbrace \mathbf{c}_{x}=\bigl(\mathrm{Tr}(xd_{1}),\mathrm{Tr}(xd_{2}),\cdots,\mathrm{Tr}(xd_{n})\bigr):x\in \mathbb{F}_{q} \biggr\rbrace.
\end{equation}
It should be mentioned that the defining set (\ref{definig_set}) is the combination of the ones of \cite{BB33} and \cite{BB35}. Note that \cite{BB33} determined just the Hamming weight enumerator of linear codes from quadratic function.

The rest of the paper is structured as follows: in Section 2, the main theorems and some examples are presented. In Section 3, mathematical foundation and proofs of the main theorems are given. Finally, some concluding remarks and the application of the linear codes constructed in the present paper are discussed in Section 4.

\section{Main theorems}
We only present the main theorems and some examples in this section. 
Two cases are treated: $ e/t \equiv 1 \pmod 2$ and $ e/t \equiv 0 \pmod 2$. 
For given $ e, t $ and $ p $, define the following constants which will be introduced in Section 3:
\begin{equation}\label{constant_kappa}
 \kappa=
 \begin{cases}
 (-1)^{e+t}p^{\frac{e-t-2}{2}} &\quad \text{if}\ p\equiv 1\pmod 4,\\
 (-1)^{t}\sqrt{-1}^{e+t}p^{\frac{e-t-2}{2}} &\quad \text{if}\ p\equiv 3\pmod 4, 
 \end{cases}
\end{equation}
and
\begin{equation}\label{constant_epsilon}
 \epsilon=
 \begin{cases}
 1 &\quad \text{if}\ p\equiv 1\pmod 4,\\
 (-1)^{e} &\quad \text{if}\ p\equiv 3\pmod 4. 
 \end{cases}
\end{equation}

\subsection{The first case: $ e/t \equiv 1 \pmod 2 $}
 \begin{thm1}\label{thm1_lab}
 Suppose that $ e/d $ is odd, $ 1<t<d, t\mid d $ and $ e/t $ is odd. Then, $ \mathcal{C}_{D_{0}} $, defined by (\ref{linear_code_cda}),  is an $ [p^{e-t}-1,e]- $ linear code. The Hamming weight distribution of $ \mathcal{C}_{D_{0}} $ is given in Table \ref{Tab1} and its complete weight enumerator is given by the following formula:
 \begin{equation}\label{cweformula_thm1}
 CWE(\mathcal{C}_{D_{0}})=w_{0}^{A_{0}}+F_{1}w_{0}^{B_{0}}\prod_{i=1}^{p-1}w_{i}^{B_{1}}+F_{2}w_{0}^{C_{0}}\prod_{i=1}^{p-1}w_{i}^{C_{1}}+F_{3}w_{0}^{D_{0}}\prod_{i=1}^{p-1}w_{i}^{D_{1}},
 \end{equation}
 where
 \[
  \begin{array}{lll}
   A_{0}=p^{e-t}-1,\\
   B_{0}=p^{e-t-1}-1, & B_{1}=p^{e-t-1}, & F_{1}=p^{e-t}-1,\\
   C_{0}=p^{e-t-1}+(p-1)\kappa-1, & C_{1}=p^{e-t-1}-\kappa, & F_{2}=\frac{1}{2}(p^{t}-1)(p^{e-t}+\epsilon\kappa\eta^{(t)}(-1)p),\\
   D_{0}=p^{e-t-1}-(p-1)\kappa-1, & D_{1}=p^{e-t-1}+\kappa, & F_{3}=\frac{1}{2}(p^{t}-1)(p^{e-t}-\epsilon\kappa\eta^{(t)}(-1)p).
   \end{array}
 \] 
\end{thm1}
 \begin{table}[tbh]
   \caption{Weight distribution of the code $ \mathcal{C}_{D_{0}} $ of Theorem \ref{thm1_lab}} 
   \label{Tab1}
   {\renewcommand{\tabcolsep}{0.15cm}
   \begin{center}
   \begin{tabular}{|l|l|}
   \hline
   weight $ w $ & multiplicity $ F_{w} $ \\
   \hline
   0 & 1\\
   \hline
   $ (p-1)p^{e-t-1} $ & $ p^{e-t}-1 $\\
   \hline
   $ (p-1)(p^{e-t-1}-\kappa) $ & $ \frac{1}{2}(p^{t}-1)(p^{e-t}+\epsilon\kappa\eta^{(t)}(-1)p) $\\
   \hline
   $ (p-1)(p^{e-t-1}+\kappa) $ & $ \frac{1}{2}(p^{t}-1)(p^{e-t}-\epsilon\kappa\eta^{(t)}(-1)p) $\\
   \hline
   \end{tabular}
   \end{center}
   }
   \end{table} 
The following example demonstrates Theorem \ref{thm1_lab}:
\begin{expn1}
Let $ p=3,e=6,\alpha=6,t=2 $. It is clear that the condition of Theorem \ref{thm1_lab} is fulfilled  with these parameters. The linear code $ \mathcal{C}_{D_{0}} $, defined by (\ref{linear_code_cda}), is an $ [80,6,48]-$ linear code. The Hamming weight enumerator is $ 1+360z^{48}+80z^{54}+288z^{60} $, and the complete weight enumerator is $ w_{0}^{80}+360w_{0}^{32}w_{1}^{24}w_{2}^{24}+80w_{0}^{26}w_{1}^{27}w_{2}^{27}+288w_{0}^{20}w_{1}^{30}w_{2}^{30} $, which are confirmed by a computer program.
\end{expn1}     
   The following theorem determines the parameters and the complete weight enumerators of $\mathcal{C}_{D_{a}} $, where $ a\in \mathbb{F}_{p^{t}}^{*} $:
    \begin{thm2}\label{thm2_lab}
    Suppose that $ e/d $ is odd, $ 1<t<d, t\mid d $ and $ e/t $ is odd. Then, the code $ \mathcal{C}_{D_{a}} $, where $ a\in \mathbb{F}_{p^{t}}^{*} $, is an $ [p^{e-t}+\epsilon\kappa\eta^{(t)}(-a)p,e]- $ linear code. The Hamming weight distribution of $ \mathcal{C}_{D_{a}} $, defined by (\ref{linear_code_cda}),  is given in Table \ref{Tab2} and its complete weight enumerator is given by the following formula:
    \begin{equation}\label{cweformula_thm3}
    \begin{split}
    CWE(\mathcal{C}_{D_{0}})&=w_{0}^{A_{0}}+F_{1}w_{0}^{B_{0}}\prod_{i=1}^{p-1}w_{i}^{B_{1}}+F_{2}w_{0}^{C_{0}}\prod_{i=1}^{p-1}w_{i}^{C_{1}}\\
       &+F_{3}w_{0}^{D_{0}}\prod_{i=1}^{p-1}w_{i}^{D_{1}}+F_{4}w_{0}^{E_{1}}\sum_{i=1}^{p-1}w_{\pm 2i\pmod p}^{E_{0}}\prod_{\substack{1\leq j\leq p-1\\j\ne \pm 2i\pmod p}}w_{j}^{E_{1}},
    \end{split}
    \end{equation}
    where
    \[
     \begin{array}{ll}
      A_{0}=p^{e-t}+\epsilon\kappa\eta^{(t)}(-a)p,\\
      B_{0}=p^{e-t-1}+\kappa(\epsilon\eta^{(t)}(-a)+\eta^{(t)}(a)(p-1)), & B_{1}=p^{e-t-1}+\kappa(\epsilon\eta^{(t)}(-a)-\eta^{(t)}(a)),\\
      C_{0}=p^{e-t-1}+\epsilon\kappa\eta^{(t)}(-a), & C_{1}=p^{e-t-1}+\epsilon\kappa\eta^{(t)}(-a),\\
      D_{0}=p^{e-t-1}+\kappa(\epsilon\eta^{(t)}(-a)+2\eta^{(t)}(a)(p-1)), & D_{1}=p^{e-t-1}+\kappa(\epsilon\eta^{(t)}(-a)-2\eta^{(t)}(a)),\\
      E_{0}=p^{e-t-1}+\kappa(\epsilon\eta^{(t)}(-a)+\eta^{(t)}(a)(p-2)),&
      E_{1}=p^{e-t-1}+\kappa(\epsilon\eta^{(t)}(-a)-2\eta^{(t)}(a)),\\
      F_{1}=p^{e-t}-1,& F_{2}=\frac{1}{2}(p^{t}-1)(p^{e-t}-\epsilon\kappa\eta^{(t)}(-a)p),\\
      F_{3}=\frac{1}{2}(p^{t-1}-1)(p^{e-t}+\epsilon\kappa\eta^{(t)}(-a)p),& F_{4}=\frac{1}{2}p^{t-1}(p^{e-t}+\epsilon\kappa\eta^{(t)}(-a)p).
      \end{array}
    \] 
   \end{thm2}
   \begin{table}[tbh]
      \caption{Weight distribution of the code $ \mathcal{C}_{D_{a}} $ of Theorem \ref{thm2_lab}} 
      \label{Tab2}
      {\renewcommand{\tabcolsep}{0.15cm}
      \begin{center}
      \begin{tabular}{|l|l|}
      \hline
      weight $ w $ & multiplicity $ F_{w} $ \\
      \hline
      0 & 1\\
      \hline
      $ (p-1)(p^{e-t-1}+\kappa(\epsilon\eta^{(t)}(-a)-\eta^{(t)}(a)) $ & $ p^{e-t}-1 $\\
      \hline
      $ (p-1)(p^{e-t-1}+\kappa\epsilon\eta^{(t)}(-a)) $ & $ \frac{1}{2}(p^{t}-1)(p^{e-t}-\epsilon\kappa\eta^{(t)}(-a)p) $\\
      \hline
      $ (p-1)(p^{e-t-1}+\kappa(\epsilon\eta^{(t)}(-a)-2\eta^{(t)}(a)) $ & $ \frac{1}{2}(p^{t-1}-1)(p^{e-t}+\epsilon\kappa\eta^{(t)}(-a)p) $\\
      \hline
      $ (p-1)(p^{e-t-1}+\kappa(\epsilon\eta^{(t)}(-a)-2\eta^{(t)}(a))+2\kappa\eta^{(t)}(a)p $ & $ \frac{1}{2}(p-1)p^{t-1}(p^{e-t}+\epsilon\kappa\eta^{(t)}(-a)p) $\\
      \hline
      \end{tabular}
      \end{center}
      }
      \end{table} 
       
    The following example is calculated from the formula (\ref{cweformula_thm3}) of Theorem \ref{thm2_lab} and Table \ref{Tab2}, and is confirmed by a computer program:
    \begin{expn2}
    Let $ p=3, e=6, \alpha=6, t=2 $. In addition, set $ a=1 $ in the defining set $ D_{a} $ of (\ref{definig_set}). Then, the code $ \mathcal{C}_{D_{1}} $, defined by (\ref{linear_code_cda}), is an $ [90,6,48]-$ linear code, with $ 1+90z^{48}+80z^{54}+288z^{60}+270z^{66} $ as the Hamming weight enumerator, and 
    \begin{equation*}
     w_{0}^{90}+90w_{0}^{42}w_{1}^{24}w_{2}^{24}+80w_{0}^{36}w_{1}^{27}w_{2}^{27}+288w_{0}^{30}w_{1}^{30}w_{2}^{30}+270w_{0}^{24}w_{1}^{33}w_{2}^{33} 
    \end{equation*}
     as its complete weight enumerator.
    \end{expn2}
   \subsection{The second case: $ e/t \equiv 0 \pmod 2 $}
   For the second case that $ e/t\equiv 0 \pmod 2 $, we only give the parameters and the complete weight enumerators of the codes $ \mathcal{C}_{D_{0}} $. We are not able to determine the ones of the codes $ \mathcal{C}_{D_{a}} $  where $ a\ne 0 $ because for this sub-case, a particular Kloosterman sum is involved and its explicit evaluation is intractable. We need some constant $ \kappa_{2} $, which occurs in \cite[Theorem 1]{BB10} and defined later by (\ref{constant_kappa_2}), to present the following theorem:
    \begin{thm3}\label{thm3_lab}
    Suppose that $ e/d $ is odd, $ 1<t<d, t\mid d $ and $ e/t $ is even. Then, $ \mathcal{C}_{D_{0}} $ is an $ [p^{e-t}+\epsilon\kappa_{2} (p^{t}-1)/p^{t}-1,e]- $ linear code. The Hamming weight distribution of $ \mathcal{C}_{D_{0}} $  is given in Table \ref{Tab3} and its complete weight enumerator is given by the following formula:
    \begin{equation}\label{cweformula_thm5}
    CWE(\mathcal{C}_{D_{0}})=w_{0}^{A_{0}}+F_{1}w_{0}^{B_{0}}\prod_{i=1}^{p-1}w_{i}^{B_{1}}+F_{2}w_{0}^{C_{0}}\prod_{i=1}^{p-1}w_{i}^{C_{1}},
    \end{equation}
    where
    \[
     \begin{array}{lll}
      A_{0}=p^{e-t}+\epsilon\kappa p(p^{t}-1)-1,\\
      B_{0}=p^{e-t-1}+\kappa(p+\epsilon-1)(p^{t}-1)-1, & B_{1}=p^{e-t-1}+\kappa(\epsilon-1)(p^{t}-1),\\
      C_{0}=p^{e-t-1}+\kappa(\epsilon(p^{t}-1)-p+1)-1, & C_{1}=p^{e-t-1}+\kappa(\epsilon(p^{t}-1)+1),\\
      F_{1}=p^{e-t}+\epsilon\kappa p(p^{t}-1)-1, &  F_{2}=(p^{t}-1)(p^{e-t}-\epsilon\kappa p).
      \end{array}
    \] 
   \end{thm3}
    \begin{table}[tbh]
      \caption{Weight distribution of the code $ \mathcal{C}_{D_{0}} $ of Theorem \ref{thm3_lab}} 
      \label{Tab3}
      {\renewcommand{\tabcolsep}{0.15cm}
      \begin{center}
      \begin{tabular}{|l|l|}
      \hline
      weight $ w $ & multiplicity $ F_{w} $ \\
      \hline
      0 & 1\\
      \hline
      $ (p-1)(p^{e-t-1}+\kappa(\epsilon(p^{t}-1)+1)) $ & $ (p^{t}-1)(p^{e-t}-\epsilon\kappa p) $\\
      \hline
      $ (p-1)(p^{e-t-1}+\kappa(\epsilon-1)(p^{t}-1)) $ & $ p^{e-t}+\epsilon\kappa p(p^{t}-1)-1 $\\
      \hline
      \end{tabular}
      \end{center}
      }
      \end{table} 
                    
The following example is calculated from the formula (\ref{cweformula_thm5}) of Theorem \ref{thm3_lab} and Table \ref{Tab3}, and confirmed by a computer program:
\begin{expn3}
Let $ p=3, e=8, \alpha=8, t=2 $. It is clear that the condition of Theorem \ref{thm3_lab} is satisfied. Then, the code $ \mathcal{C}_{D_{0}} $, defined by (\ref{linear_code_cda}), is an $ [656,8,432]-$ linear code. The Hamming weight enumerator is $ 1+5904z^{432}+656z^{486} $, and the complete weight enumerator is $ w_{0}^{656}+5904w_{0}^{224}w_{1}^{216}w_{2}^{216}+656w_{0}^{170}w_{1}^{243}w_{2}^{243} $.
\end{expn3}

   \section{Mathematical foundation and proofs of the main theorems}
   
   An \textit{additive character} of $ \mathbb{F}_{p^{n}} $ is a nonzero function $ \chi $ from $ \mathbb{F}_{p^{n}} $ to a set of nonzero complex numbers of absolute value 1 such that for all $ x,y\in  \mathbb{F}_{p^{n}}  $, $ \chi(x+y)=\chi(x)\chi(y) $. Let $ \zeta_{p}=e^{2\pi\sqrt{-1}/p} $. For each $ b\in \mathbb{F}_{p^{n}}$, the function 
   \begin{equation*}
   \chi_{b}(x)=\zeta_{p}^{\mathrm{Tr}(bx)}\ \text{for all}\ x\in \mathbb{F}_{p^{n}},
   \end{equation*}
   defines an additive character of $ \mathbb{F}_{p^{n}} $. The character $ \chi_{0} $ is called the \textit{trivial additive character} of $ \mathbb{F}_{p^{n}} $, and $ \chi_{1} $ is called the \textit{canonical additive character} of $ \mathbb{F}_{p^{n}} $. In this paper, the subscript of the canonical additive character of $  \mathbb{F}_{p^{n}} $ is  omitted, and the canonical additive character of $ \mathbb{F}_{p^{k}} $ is denoted by $ \chi^{(k)} $, where $ k $ is a positive integer.
   
   An \textit{multiplicative character} of $ \mathbb{F}_{p^{n}} $ is a nonzero function $ \psi $ from $ \mathbb{F}_{p^{n}}^{*} $ to a set of nonzero complex numbers of absolute value 1 such that for all $ x,y\in  \mathbb{F}_{p^{n}}^{*}  $, $ \psi(xy)=\psi(x)\psi(y) $. Let $ \theta $ be a primitive element of $ \mathbb{F}_{p^{n}}^{*} $. Then, all the multiplicative characters are given by
   \begin{equation*}
   \psi_{j}(\theta^{k})=e^{2\pi\sqrt{-1}jk/(p^{n}-1)},0\leq k\leq p^{n}-2,
   \end{equation*}
   for  $ 0\leq j\leq p^{n}-2 $. The multiplicative character $ \psi_{(p^{n}-1)/2} $ is called the \textit{quadratic character} of $ \mathbb{F}_{p^{n}} $. Denote the quadratic character of an arbitrary field $ \mathbb{F}_{p^{k}} $ by $ \eta^{(k)} $.
   
   The quadratic Gauss sum over  $ \mathbb{F}_{p^{k}} $ is defined by
   \begin{equation*}
   G(\eta^{(k)})=\sum_{x\in \mathbb{F}_{p^{k}}^{*}}\eta^{(k)}(x)\chi^{(k)}(x).
   \end{equation*}
   
   \begin{lemma1}\cite[Theorem 5.15]{BB27}\label{lma1_ref}
   With the notations and definitions above, we have
   \begin{equation*}
   G(\eta^{(k)})=(-1)^{k-1}\sqrt{-1}^{\frac{(p-1)^{2}k}{4}}\sqrt{p^{k}}.
   \end{equation*}
   \end{lemma1}
   
   \begin{lemma2}\cite[Theorem 5.33]{BB27}\label{lma2_ref}
   Let $ \chi^{(k)} $ be a nontrivial character of $ \mathbb{F}_{p^{k}} $, and let $ f(x)=a_{2}x^{2}+a_{1}x+a_{0} $ with $ a_{2} \ne 0$. Then
   \begin{equation*}
   \sum_{x\in \mathbb{F}_{p^{k}}}\chi^{(k)}(f(x))=\chi^{(k)}(a_{0}-a_{1}^{2}(4a_{2})^{-1})\eta^{(k)}(a_{2})G(\eta^{(k)}).
   \end{equation*}
   \end{lemma2}
   
   Let $ F(x)=ax^{p^{\alpha}+1}+bx $ be a mapping from $ \mathbb{F}_{q} $ onto itself where $ a\in \mathbb{F}_{q}^{*},b\in \mathbb{F}_{q} $. Recall that $ q=p^{e},d=\gcd(e,\alpha) $ and $ e/d $ is assumed to be odd. Define two kinds of Weil sums as follows.
   \begin{equation*}
   \begin{split}
   S_{\alpha}(a)&=\sum_{x\in \mathbb{F}_{q}}\chi^{(e)}(ax^{p^{\alpha}+1}),\\
   S_{\alpha}(a,b)&=\sum_{x\in \mathbb{F}_{q}}\chi^{(e)}(ax^{p^{\alpha}+1}+bx).
   \end{split}
   \end{equation*}
   The explicit evaluation of $ S_{\alpha}(a) $ and $ S_{\alpha}(a,b) $ are given by the following two lemmas, respectively.
   \begin{lemma3}\cite[Theorem 1]{BB09}\label{lma3_ref}
   Let $ e/d $ be odd. Then
   \begin{equation*}
   S_{\alpha}(a)=\kappa_{1}\eta^{(e)}(a),
   \end{equation*}
   where
   \begin{equation}\label{constant_kappa_1}
   \kappa_{1}=
   \begin{cases}
   (-1)^{e-1}\sqrt{q}&\quad \text{if}\ p\equiv 1\pmod 4\\
   (-1)^{e-1}\sqrt{-1}^{e}\sqrt{q}&\quad \text{if}\ p\equiv 3\pmod 4.
   \end{cases}
   \end{equation}
   \end{lemma3}
   \begin{lemma4}\cite[Theorem 1]{BB10}\label{lma4_ref}
   Let $ q=p^{e} $ where $ p $ is an odd prime, and $ e $ is a positive integer. For $ a\in \mathbb{F}^{*}_{q} $, define the function $ f(X)=a^{p^{\alpha}}X^{p^{2\alpha}} +aX$ from $ \mathbb{F}_{q} $ to itself. Let $ d=\gcd(e,\alpha) $, and suppose that $ e/d $ is odd. Then, $ f(X) $ is a permutation polynomial over $ \mathbb{F}_{q} $. Let $ \gamma $ be the unique solution of the equation $ f(X)=- b^{p^{\alpha}},b\ne 0$. We have
          \begin{equation*}
           S_{\alpha}(a,b)=\kappa_{2}\eta^{(e)}(-a)\overline{\chi^{(e)}\bigl(a\gamma^{p^{\alpha}+1}}\bigr),
           \end{equation*}
           where 
           \begin{equation}\label{constant_kappa_2}
           \kappa_{2}=
           \begin{cases}
             (-1)^{e-1}\sqrt{q} &\quad\text{if}\ p\equiv 1\pmod 4\\
             (-1)^{e-1}\sqrt{-1}^{3e}\sqrt{q}&\quad\text{if}\ p\equiv 3\pmod 4.
           \end{cases}
           \end{equation}
          \end{lemma4}
   From (\ref{constant_epsilon}), (\ref{constant_kappa_1}) and (\ref{constant_kappa_2}),  it is obvious that $ \kappa_{2}=\epsilon\kappa_{1} $ and $ \kappa_{1}=\epsilon\kappa_{2} $.
   Recall that $ q=p^{e},d=\gcd(\alpha,e), 1<t<d, t\mid d $. Define 
      \begin{equation}\label{constant_kappa_general}
      \kappa:=\kappa_{2}G(\eta^{(t)})p^{-t-1}.
      \end{equation}
      By Lemma \ref{lma1_ref} and (\ref{constant_kappa_2}), we can obtain the expanded formula for (\ref{constant_kappa_general}) which is identical to (\ref{constant_kappa}). 
   
   Let $ \mathrm{Re}(.) $ denote the real part of a complex number. The explicit evaluation of the Sali\'{e} sums defined by (\ref{TwistedKSum_def}) is given by next lemma:
   \begin{lemma5}\cite[Lemma 12.4]{BB21}, \cite[Theorem 2.19]{BB22}\label{lma5_ref}
   Let $ a,b\in \mathbb{F}_{p^{t}}^{*} $. Then
   \begin{equation*}
   T(\eta^{(t)};a,b)=
   \begin{cases}
   0 &\quad \text{if}\ \eta^{(t)}(a)\ne \eta^{(t)}(b)\\
   2\eta^{(t)}(a)G(\eta^{(t)})\mathrm{Re}\bigl(\chi^{(t)}(2\sqrt{ab})\bigr) &\quad \text{if}\ \eta^{(t)}(a)= \eta^{(t)}(b).
   \end{cases}
   \end{equation*}
   \end{lemma5}
   
   The following lemma is useful:
   \begin{lemma6}\cite[Lemma 7]{BB17}\label{lma6_ref}
   Let $ m,n>1 $ be two positive integers with $ m $ dividing $ n $. Then, for any $ x\in \mathbb{F}_{p^{m}}^{*} $, $ \eta^{(n)}(x)=1 $ if $ n/m $ is even, and $ \eta^{(n)}(x)=\eta^{(m)}(x) $ if $ n/m $ is odd.
   \end{lemma6}
   
   If $ e/d $ is odd, then $ f(x)=x^{p^{2\alpha}}+x $ is a permutation polynomial over $ \mathbb{F}_{q} $. For $ b\in \mathbb{F}_{q} $, let $ \gamma $ denote the unique solution of $ f(x)=-b^{p^{\alpha}} $. In order to simplify the notations, put $ \Delta:=\mathrm{Tr}^{e}_{t}(\gamma^{p^{\alpha}+1}) $.
   The following lemma gives the explicit evaluation of the exponential sum $ L_{\alpha,b}(a,u) $ defined by (\ref{BSum_def}), where $ a\in \mathbb{F}_{p^{t}} $ and $ u\in \mathbb{F}_{p} $ (Recall that $ q=p^{e}, d=\gcd(\alpha,e), 1<t<d, t\mid d $):
   
   \begin{lemma7}\label{lma7_ref}
   \begin{enumerate}[(I)]
   \item The first case: $ e/t\equiv 1 \pmod 2 $.
   \begin{enumerate}[(1)]
   \item $ a=0, u=0 $.
   \begin{equation*}
   L_{\alpha,b}(0,0)=
   \begin{cases}
   0 &\quad \text{if}\ \Delta =0\\
   (p-1)\kappa_{2}G(\eta^{(t)}) &\quad \text{if}\ \Delta \ne 0\ \text{and}\ \eta^{(t)}(\Delta)=1\\
   -(p-1)\kappa_{2}G(\eta^{(t)}) &\quad \text{if}\ \Delta \ne 0\ \text{and}\ \eta^{(t)}(\Delta)=-1.
   \end{cases}
   \end{equation*}
   \item $ a=0, u\in \mathbb{F}_{p}^{*} $.
    \begin{equation*}
      L_{\alpha,b}(0,u)=
      \begin{cases}
      0 &\quad \text{if}\ \Delta =0\\
      -\kappa_{2}G(\eta^{(t)}) &\quad \text{if}\ \Delta \ne 0\ \text{and}\ \eta^{(t)}(\Delta)=1\\
       \kappa_{2}G(\eta^{(t)}) &\quad \text{if}\ \Delta \ne 0\ \text{and}\ \eta^{(t)}(\Delta)=-1.
      \end{cases}
      \end{equation*}
   \item $ a\in \mathbb{F}_{p^{t}}^{*},u=0 $. Let $ L:= L_{\alpha,b}(a,0)$.
    \begin{equation*}
      L=
      \begin{cases}
      (p-1)\eta^{(t)}(a)\kappa_{2}G(\eta^{(t)}) &\quad \text{if}\ \Delta =0\\
      0 &\quad \text{if}\ \Delta \ne 0\ \text{and}\ \eta^{(t)}(\Delta)\ne \eta^{(t)}(a)\\
      2(p-1)\eta^{(t)}(a)\kappa_{2}G(\eta^{(t)})&\quad \text{if}\ \Delta \ne 0\ \text{and}\ \eta^{(t)}(\Delta)= \eta^{(t)}(a)\ \text{and} \ \mathrm{Tr}^{t}_{1}(\sqrt{a\Delta})=0\\
      -2\eta^{(t)}(a)\kappa_{2}G(\eta^{(t)})&\quad \text{if}\ \Delta \ne 0\ \text{and}\ \eta^{(t)}(\Delta)= \eta^{(t)}(a)\ \text{and} \ \mathrm{Tr}^{t}_{1}(\sqrt{a\Delta})\ne 0.
      \end{cases}
      \end{equation*}
   \item $ a\in \mathbb{F}_{p^{t}}^{*},u\in \mathbb{F}_{p}^{*} $. Let $ L:= L_{\alpha,b}(a,u)$.
    \begin{equation*}
         L=
         \begin{cases}
         -\eta^{(t)}(a)\kappa_{2}G(\eta^{(t)}) &\quad \text{if}\ \Delta =0\\
         0 &\quad \text{if}\ \Delta \ne 0\ \text{and}\ \eta^{(t)}(\Delta)\ne \eta^{(t)}(a)\\
         -2\eta^{(t)}(a)\kappa_{2}G(\eta^{(t)})&\quad \text{if}\ \Delta \ne 0\ \text{and}\ \eta^{(t)}(\Delta)= \eta^{(t)}(a)\ \text{and} \ \mathrm{Tr}^{t}_{1}(\sqrt{a\Delta})=0\\
         -2\eta^{(t)}(a)\kappa_{2}G(\eta^{(t)})&\quad \text{if}\ \Delta \ne 0\ \text{and}\ \eta^{(t)}(\Delta)= \eta^{(t)}(a)\ \text{and} \  \mathrm{Tr}^{t}_{1}(\sqrt{a\Delta})\notin \lbrace 0,\pm 2^{-1}u\rbrace\\
         (p-2)\eta^{(t)}(a)\kappa_{2}G(\eta^{(t)})&\quad \text{if}\ \Delta \ne 0\ \text{and}\ \eta^{(t)}(\Delta)= \eta^{(t)}(a)\ \text{and} \  \mathrm{Tr}^{t}_{1}(\sqrt{a\Delta})=\pm 2^{-1}u.
         \end{cases}
         \end{equation*}
   \end{enumerate}
   \item The second case: $ e/t\equiv 0 \pmod 2 $.
     \begin{enumerate}[(1)]
      \item $ a=0, u=0 $.
      \begin{equation*}
      L_{\alpha,b}(0,0)=
      \begin{cases}
      \kappa_{2}(p-1)(p^{t}-1) &\quad \text{if}\ \Delta =0\\
     -\kappa_{2}(p-1)          &\quad \text{if}\ \Delta \ne 0.
      \end{cases}
      \end{equation*}
      \item $ a=0, u\in \mathbb{F}_{p}^{*} $.
       \begin{equation*}
         L_{\alpha,b}(0,u)=
         \begin{cases}
        -\kappa_{2}(p^{t}-1) &\quad \text{if}\ \Delta =0\\
        \kappa_{2}           &\quad \text{if}\ \Delta \ne 0.
         \end{cases}
         \end{equation*}
       \end{enumerate}
   \end{enumerate}
   \end{lemma7}
   \begin{proof}
   \begin{enumerate}[(I)]
   \item The first case: $ e/t\equiv 1 \pmod 2 $.\newline For this case, we only give the proof of the sub-case (4) since the proofs for the remainder sub-cases are very similar.  Let $ y\in \mathbb{F}_{p^{t}}^{*}, z\in \mathbb{F}_{p}^{*} $. According to Lemma \ref{lma6_ref}, $ \eta^{(e)}(y)=\eta^{(t)}(y) $ since $ e/t $ is odd. For a given $ b\in \mathbb{F}_{q}^{*} $, let $ \gamma $ be the unique solution of the equation $ x^{p^{2\alpha}}+x+b^{p^{\alpha}}=0 $ over $ \mathbb{F}_{q} $. It is easy to check that $ y^{-1}\gamma z $ is a solution of the equation $ yx^{p^{2\alpha}}+yx+(zb)^{p^{\alpha}}=0 $.  From (\ref{BSum_def}) and  Lemma \ref{lma4_ref}, we have
   \begin{equation}\label{proof_BSum}
   \begin{split}
   L_{\alpha,b}(a,u)&=\sum_{z\in \mathbb{F}_{p}^{*}}\chi^{(1)}(-uz)\sum_{y\in \mathbb{F}_{p^{t}}^{*}}\chi^{(t)}(-ay)\sum_{x\in \mathbb{F}_{q}}\chi^{(e)}(yx^{p^{\alpha}+1}+zbx)\\
   &=\sum_{z\in \mathbb{F}_{p}^{*}}\chi^{(1)}(-uz)\sum_{y\in \mathbb{F}_{p^{t}}^{*}}\chi^{(t)}(-ay)S_{\alpha}(y,zb)\\
   &=\sum_{z\in \mathbb{F}_{p}^{*}}\chi^{(1)}(-uz)\sum_{y\in \mathbb{F}_{p^{t}}^{*}}\chi^{(t)}(-ay)\kappa_{2}\eta^{(e)}(-y)\overline{\chi^{e}\bigl(y(y^{-1}\gamma z)^{p^{\alpha}+1}\bigr)}\\
   &=\sum_{z\in \mathbb{F}_{p}^{*}}\chi^{(1)}(-uz)\sum_{y\in \mathbb{F}_{p^{t}}^{*}}\chi^{(t)}(-ay)\kappa_{2}\eta^{(t)}(-y)\chi^{(t)}\bigl(-y^{-1}\mathrm{Tr}^{e}_{t}(z^{2}\gamma^{p^{\alpha}+1})\bigr)\\
   &=\kappa_{2}\sum_{z\in \mathbb{F}_{p}^{*}}\chi^{(1)}(uz)T\bigl(\eta^{(t)};\mathrm{Tr}^{e}_{t}(z^{2}\gamma^{p^{\alpha}+1}),a\bigr),
   \end{split}
   \end{equation}
   where 
   \begin{equation*}
   T(\eta^{(t)};\mathrm{Tr}^{e}_{t}(z^{2}\gamma^{p^{\alpha}+1}),a)=\sum_{y\in \mathbb{F}_{p^{t}}^{*}}\eta^{(t)}(y)\chi^{(t)}\bigl(ay+\mathrm{Tr}^{e}_{t}(z^{2}\gamma^{p^{\alpha}+1})y^{-1}\bigr)
   \end{equation*}
   is the twisted Kloosterman sum defined by (\ref{TwistedKSum_def}). 
   \begin{enumerate}[(a)]
   \item If $ \Delta:=\mathrm{Tr}^{e}_{t}(\gamma^{p^{\alpha}+1})=0 $, then, by Lemma \ref{lma5_ref}, we have
   \begin{equation*}
   \begin{split}
   T(\eta^{(t)};0,a)&=\sum_{y\in \mathbb{F}_{p^{t}}^{*}}\eta^{(t)}(y)\chi^{(t)}(ay)=\eta^{(t)}(a)G(\eta^{(t)}),\\
   L_{\alpha,b}(a,u)&=\kappa_{2}\sum_{z\in \mathbb{F}_{p}^{*}}\chi^{(1)}(uz)T\bigl(\eta^{(t)};0,a\bigr)\\
   &=\kappa_{2}G(\eta^{(t)})\eta^{(t)}(a)\sum_{z\in \mathbb{F}_{p}^{*}}\chi^{(1)}(uz)\\
   &=-\eta^{(t)}(a)\kappa_{2}G(\eta^{(t)}).
   \end{split}
   \end{equation*}
   \item If $ \Delta\ne 0 $ and $ \eta^{(t)}(a)\ne \eta^{(t)}(\Delta) $, then, by Lemma \ref{lma5_ref},  $ T(\eta^{(t)};z^{2}\Delta,a)=0 $, which leads to that $ L_{\alpha,b}(a,u)=0 $.  
   \item If $ \Delta\ne 0 $ and $ \eta^{(t)}(a)= \eta^{(t)}(\Delta) $, then, by Lemma \ref{lma5_ref}, 
   \begin{equation*}
   \begin{split}
   L_{\alpha,b}(a,u)&=\kappa_{2}\sum_{z\in \mathbb{F}_{p}^{*}}\chi^{(1)}(uz)T\bigl(\eta^{(t)};z^{2}\Delta,a\bigr)\\
   &=\kappa_{2}\sum_{z\in \mathbb{F}_{p}^{*}}\chi^{(1)}(uz)\eta^{(t)}(z^{2}\Delta)G(\eta^{(t)})\bigl(\chi^{(t)}(2\sqrt{a\Delta}z)+\overline{\chi^{(t)}(2\sqrt{a\Delta}z)}\bigr)\\
   &=\eta^{(t)}(a)\kappa_{2}G(\eta^{(t)})\biggl(\sum_{z\in \mathbb{F}_{p}^{*}}\zeta_{p}^{(u+2\mathrm{Tr}^{t}_{1}(\sqrt{a\Delta}))z}+\sum_{z\in \mathbb{F}_{p}^{*}}\zeta_{p}^{(u-2\mathrm{Tr}^{t}_{1}(\sqrt{a\Delta}))z}\biggr).
   \end{split}
   \end{equation*} 
   For the rest of analysis, we distinguish three cases: $\mathrm{Tr}^{t}_{1}(\sqrt{a\Delta}))=0,\mathrm{Tr}^{t}_{1}(\sqrt{a\Delta}))\notin \lbrace 0,\pm 2^{-1}u\rbrace $ and $ \mathrm{Tr}^{t}_{1}(\sqrt{a\Delta}))=\pm 2^{-1}u $. The further analysis is straightforward and omitted.                                 
   \end{enumerate}
   \item The second case: $ e/t\equiv 0 \pmod 2 $. \newline By Lemma \ref{lma6_ref}, $ \eta^{(e)}(y)=1 $ for all $ y\in \mathbb{F}_{p^{t}}^{*} $. Based on the analysis of the first case that $ e/t\equiv 1 \pmod 2 $ and the third equality of (\ref{proof_BSum}), we obtain
   \begin{equation*}
   \begin{split}
    L_{\alpha,b}(a,u)&=\sum_{z\in \mathbb{F}_{p}^{*}}\chi^{(1)}(-uz)\sum_{y\in \mathbb{F}_{p^{t}}^{*}}\chi^{(t)}(-ay)\kappa_{2}\eta^{(e)}(-y)\overline{\chi^{e}\bigl(y(y^{-1}\gamma z)^{p^{\alpha}+1}\bigr)}\\
    &=\kappa_{2}\sum_{z\in \mathbb{F}_{p}^{*}}\chi^{(1)}(uz)\sum_{y\in \mathbb{F}_{p^{t}}^{*}}\chi^{(t)}\biggl(ay+\mathrm{Tr}^{e}_{t}\bigl(z^{2}\gamma^{p^{\alpha}+1}\bigr)y^{-1}\biggr)\\
    &=\kappa_{2}\sum_{z\in \mathbb{F}_{p}^{*}}\chi^{(1)}(uz)\sum_{y\in \mathbb{F}_{p^{t}}^{*}}\chi^{(t)}\biggl(ay+z^{2}\Delta y^{-1}\biggr)\\
    &=\kappa_{2}\sum_{z\in \mathbb{F}_{p}^{*}}\chi^{(1)}(uz)K(\eta^{(t)};z^{2}\Delta,a),
   \end{split}
   \end{equation*} 
   where 
   \begin{equation}\label{kloosterman_sum}
   K(\eta^{(t)};z^{2}\Delta,a)=\sum_{y\in \mathbb{F}_{p^{t}}^{*}}\chi^{(t)}\biggl(ay+z^{2}\Delta y^{-1}\biggr)
   \end{equation} 
   is the Kloosterman sum defined by (\ref{KSum_def}). If $ a\ne 0 $ and $ \Delta:= \mathrm{Tr}^{e}_{t}\bigl(\gamma^{p^{\alpha}+1}\bigr)\ne 0$, it is difficult to obtain the explicit evaluation of the related Kloosterman sum. Hence, we only consider the case that $ a=0 $. Thus,
   \begin{equation*}
   \begin{split}
   L_{\alpha,b}(0,u)&=\kappa_{2}\sum_{z\in \mathbb{F}_{p}^{*}}\chi^{(1)}(uz)\sum_{y\in \mathbb{F}_{p^{t}}^{*}}\chi^{(t)}\biggl(z^{2}\Delta y^{-1}\biggr)\\
   &=\kappa_{2}\sum_{z\in \mathbb{F}_{p}^{*}}\chi^{(1)}(uz)\sum_{y\in \mathbb{F}_{p^{t}}^{*}}\chi^{(t)}\biggl(z^{2}\Delta y\biggr).
   \end{split}
   \end{equation*} 
   Further analysis distinguishes two cases: $ \Delta=0 $ and $ \Delta\ne 0 $, of which we omit the details.
   \end{enumerate}
   The proof is completed.
   \end{proof}
   
   Let $ a\in \mathbb{F}_{p^{t}} $. Define the exponential sum
   \begin{equation}\label{MSum_ref}
   M_{\alpha}(a)=\sum_{y\in \mathbb{F}_{p^{t}}^{*}}\chi^{(t)}\biggl(-ay\biggr)\sum_{x\in \mathbb{F}_{q}}\chi^{(e)}\biggl(yx^{p^{\alpha}+1}\biggr).
   \end{equation}
   Next lemma gives the explicit evaluation of (\ref{MSum_ref}).
   \begin{lemma8}\label{lma8_ref}
   \begin{enumerate}[(I)]
   \item The first case: $ e/t\equiv 1\pmod 2 $.
   \begin{equation*}
    M_{\alpha}(a)=
    \begin{cases}
    0 &\quad \text{if}\ a=0\\
    \eta^{(t)}(-a)\epsilon\kappa_{2}G(\eta^{(t)}) &\quad \text{if}\ a\ne 0.
    \end{cases}
   \end{equation*}
   \item The second case: $ e/t\equiv 0\pmod 2 $.
      \begin{equation*}
       M_{\alpha}(a)=
       \begin{cases}
       \epsilon\kappa_{2}(p^{t}-1) &\quad \text{if}\ a=0\\
       -\epsilon\kappa_{2} &\quad \text{if}\ a\ne 0.
       \end{cases}
      \end{equation*}
   \end{enumerate}
   \end{lemma8}
   \begin{proof}
   From Lemma \ref{lma3_ref}, Lemma \ref{lma6_ref}, (\ref{constant_kappa_1}) and (\ref{constant_kappa_2}), we have
   \begin{equation*}
   \begin{split}
    M_{\alpha}(a)&=\sum_{y\in \mathbb{F}_{p^{t}}^{*}}\chi^{(t)}\biggl(-ay\biggr)\sum_{x\in \mathbb{F}_{q}}\chi^{(e)}\biggl(yx^{p^{\alpha}+1}\biggr)
    =\sum_{y\in \mathbb{F}_{p^{t}}^{*}}\chi^{(t)}\biggl(-ay\biggr)S_{\alpha}(y)
    =\sum_{y\in \mathbb{F}_{p^{t}}^{*}}\chi^{(t)}\biggl(-ay\biggr)\kappa_{1}\eta^{(e)}(y)\\
    &=
    \begin{cases}
    \epsilon\kappa_{2}\sum_{y\in \mathbb{F}_{p^{t}}^{*}}\chi^{(t)}\biggl(-ay\biggr)\eta^{(t)}(y)&\quad \text{if}\ e/t\equiv 1 \pmod 2\\
    \epsilon\kappa_{2}\sum_{y\in \mathbb{F}_{p^{t}}^{*}}\chi^{(t)}\biggl(-ay\biggr)&\quad \text{if}\ e/t\equiv 0 \pmod 2.
    \end{cases}
   \end{split}
   \end{equation*}
   The further analysis is straightforward and omitted. The proof is completed.
   \end{proof}
   
   In order to establish the complete weight enumerators and the weight distributions of the codes $ \mathcal{C}_{D_{a}} $ defined by (\ref{linear_code_cda}), consider the following subset of $ \mathbb{F}_{q} $:
      \begin{equation}\label{NSet_ref}
     N_{\alpha,b}(a,u)=\biggl\lbrace x\in \mathbb{F}_{q}:\mathrm{Tr}^{e}_{t}\bigl(x^{p^{\alpha}+1}\bigr)=a\ \text{and}\ \mathrm{Tr}\bigl(bx\bigr)=u\biggr\rbrace,
      \end{equation}
   where $ a\in \mathbb{F}_{p^{t}}, u\in \mathbb{F}_{p} $ and $ b\in \mathbb{F}_{q}^{*} $. Let $ \mathbf{c}_{b}=(\mathrm{Tr}(bd_{1}),\mathrm{Tr}(bd_{2}),\cdots,\mathrm{Tr}(bd_{n})) $ be a codeword of the codes $ \mathcal{C}_{D_{a}} $ with the defining set $ D_{a}=(d_{1},d_{2},\cdots,d_{n}) $ defined by (\ref{definig_set}). Denote the length of codewords by 
   \begin{equation}\label{codelength_ref}
   n_{\alpha}(a)=
   \begin{cases}
   \#D_{a} & \qquad\text{if}\ a\ne 0\\
   \#D_{a}\cup \lbrace 0\rbrace & \qquad\text{if}\ a=0.
   \end{cases}
   \end{equation}
   Then, the Hamming weight of $ \mathbf{c}_{b} $ equals to
   \begin{equation*}
   wt(\mathbf{c}_{b})=n_{\alpha}(a)-\#N_{\alpha,b}(a,0).
   \end{equation*}

   Next lemma determines the cardinality of the set $ N_{\alpha,b}(a,u) $.
   
   \begin{lemma9}\label{lma9_ref} For a given $ b\in \mathbb{F}_{q}^{*} $, denote the unique solution of $ x^{p^{2\alpha}}+x+b^{p^{\alpha}}=0 $ over $ \mathbb{F}_{q} $ by $ \gamma $ and put $ \Delta:=\mathrm{Tr}^{e}_{t}(\gamma^{p^{\alpha}+1}) $.
   \begin{enumerate}[(I)]
   \item The first case: $ e/t\equiv 1\pmod 2 $.
   \begin{enumerate}[(a)]
   \item $ a=0,u=0 $.
   \begin{equation*} 
   \#N_{\alpha,b}(0,0)=
   \begin{cases}
      p^{e-t-1}-1 &\quad \text{if}\ \Delta=0\\
      p^{e-t-1}+(p-1)\kappa-1 &\quad \text{if}\ \Delta \ne 0\ \text{and}\ \eta^{(t)}(\Delta)=1\\
      p^{e-t-1}-(p-1)\kappa-1 &\quad \text{if}\ \Delta \ne 0\ \text{and}\ \eta^{(t)}(\Delta)=-1.
   \end{cases}
   \end{equation*}
   \item $ a=0,u\ne 0 $.
   \begin{equation*} 
      \#N_{\alpha,b}(0,u)=
      \begin{cases}
         p^{e-t-1}-1 &\quad \text{if}\ \Delta=0\\
         p^{e-t-1}-\kappa &\quad \text{if}\ \Delta \ne 0\ \text{and}\ \eta^{(t)}(\Delta)=1\\
         p^{e-t-1}+\kappa &\quad \text{if}\ \Delta \ne 0\ \text{and}\ \eta^{(t)}(\Delta)=-1.
      \end{cases}
      \end{equation*}
   \item $ a\ne 0,u=0 $.
   \begin{equation*} 
      \#N_{\alpha,b}(a,0)=
      \begin{cases}
       N_{1}   &\quad \text{if}\ \Delta=0\\
       N_{2}  &\quad \text{if}\ \Delta \ne 0\ \text{and}\ \eta^{(t)}(\Delta)\ne \eta^{(t)}(a)\\
       N_{3}  &\quad \text{if}\ \Delta \ne 0\ \text{and}\ \eta^{(t)}(\Delta)= \eta^{(t)}(a)\ \text{and}\ \mathrm{Tr}^{t}_{1}(\sqrt{a\Delta})=0\\
       N_{4}   &\quad \text{if}\ \Delta \ne 0\ \text{and}\ \eta^{(t)}(\Delta)= \eta^{(t)}(a)\ \text{and}\ \mathrm{Tr}^{t}_{1}(\sqrt{a\Delta})\ne 0,
      \end{cases}
      \end{equation*}
      where
             \[
              \begin{array}{ll}
              N_{1}=p^{e-t-1}+\kappa\biggl(\epsilon\eta^{(t)}(-a)+(p-1)\eta^{(t)}(a)\biggr), & N_{2}=p^{e-t-1}+\kappa\epsilon\eta^{(t)}(-a),\\
              N_{3}=p^{e-t-1}+\kappa\biggl(\epsilon\eta^{(t)}(-a)+2(p-1)\eta^{(t)}(a)\biggr), & N_{4}=p^{e-t-1}+\kappa\biggl(\epsilon\eta^{(t)}(-a)-2\eta^{(t)}(a)\biggr).
              \end{array}
             \] 
   \item $ a\ne 0,u\ne 0 $.
   \begin{equation*} 
         \#N_{\alpha,b}(a,u)=
         \begin{cases}
          N_{1}   &\quad \text{if}\ \Delta=0\\
          N_{2}  &\quad \text{if}\ \Delta \ne 0\ \text{and}\ \eta^{(t)}(\Delta)\ne \eta^{(t)}(a)\\
          N_{3}  &\quad \text{if}\ \Delta \ne 0\ \text{and}\ \eta^{(t)}(\Delta)= \eta^{(t)}(a)\ \text{and}\ \mathrm{Tr}^{t}_{1}(\sqrt{a\Delta})=0\\
          N_{3}  &\quad \text{if}\ \Delta \ne 0\ \text{and}\ \eta^{(t)}(\Delta)= \eta^{(t)}(a)\ \text{and}\ \mathrm{Tr}^{t}_{1}(\sqrt{a\Delta})\notin \lbrace 0,\pm 2^{-1}u\rbrace\\
          N_{4}   &\quad \text{if}\ \Delta \ne 0\ \text{and}\ \eta^{(t)}(\Delta)= \eta^{(t)}(a)\ \text{and}\ \mathrm{Tr}^{t}_{1}(\sqrt{a\Delta})=\pm 2^{-1}u,
         \end{cases}
         \end{equation*}
         where
                      \[
                       \begin{array}{ll}
                       N_{1}=p^{e-t-1}+\kappa\biggl(\epsilon\eta^{(t)}(-a)-\eta^{(t)}(a)\biggr), & N_{2}=p^{e-t-1}+\kappa\epsilon\eta^{(t)}(-a),\\
                       N_{3}=p^{e-t-1}+\kappa\biggl(\epsilon\eta^{(t)}(-a)-2\eta^{(t)}(a)\biggr), & N_{4}=p^{e-t-1}+\kappa\biggl(\epsilon\eta^{(t)}(-a)+(p-2)\eta^{(t)}(a)\biggr).
                       \end{array}
                      \] 
   \end{enumerate}
   \item The second case: $ e/t\equiv 0\pmod 2 $.
   \begin{enumerate}[(a)]
      \item $ a=0,u=0 $.
         \begin{equation*} 
         \#N_{\alpha,b}(0,0)=
         \begin{cases}
            p^{e-t-1}+\kappa(p+\epsilon-1)(p^{t}-1) &\quad \text{if}\ \Delta=0\\
            p^{e-t-1}+\kappa\bigl(\epsilon(p^{t}-1)-p+1\bigr) &\quad \text{if}\ \Delta \ne 0.
         \end{cases}
         \end{equation*}
         \item $ a=0,u\ne 0 $.
         \begin{equation*} 
            \#N_{\alpha,b}(0,u)=
                    \begin{cases}
                       p^{e-t-1}+\kappa(\epsilon-1)(p^{t}-1) &\quad \text{if}\ \Delta=0\\
                       p^{e-t-1}+\kappa\bigl(\epsilon(p^{t}-1)+1\bigr) &\quad \text{if}\ \Delta \ne 0.
                    \end{cases}
            \end{equation*}
      \end{enumerate}
   \end{enumerate}
   \end{lemma9}
   \begin{proof}
    By the orthogonality of additive  characters, (\ref{NSet_ref}), (\ref{BSum_def}) and  (\ref{MSum_ref}), we have
      \begin{equation*}
      \begin{split}
      \#N_{\alpha,b}(a,u)&=\sum_{x\in \mathbb{F}_{q}}p^{-t}\biggl(\sum_{y\in \mathbb{F}_{p^{t}}} \chi^{(t)}\biggl(\bigl(\mathrm{Tr}^{e}_{t}(x^{p^{\alpha}+1})-a\bigr)y\biggr)\biggr)p^{-1}\biggl(\sum_{z\in \mathbb{F}_{p}} \chi^{(1)}\biggl(\bigl(\mathrm{Tr}^{e}_{1}(bx)-u\bigr)z\biggr)\biggr)\\
      &=p^{-t-1}\sum_{x\in \mathbb{F}_{q}}\biggl(1+\sum_{y\in \mathbb{F}_{p^{t}}^{*}} \chi^{(t)}\biggl(\bigl(\mathrm{Tr}^{e}_{t}(x^{p^{\alpha}+1})-a\bigr)y\biggr)\biggr)\biggl(1+\sum_{z\in \mathbb{F}_{p}^{*}} \chi^{(1)}\biggl(\bigl(\mathrm{Tr}^{e}_{1}(bx)-u\bigr)z\biggr)\biggr)\\
      &=p^{e-t-1}+p^{-t-1}\sum_{x\in \mathbb{F}_{q}}\sum_{z\in \mathbb{F}_{p}^{*}} \chi^{(1)}\biggl(\bigl(\mathrm{Tr}^{e}_{1}(bx)-u\bigr)z\biggr)
      +p^{-t-1}\sum_{y\in \mathbb{F}_{p^{t}}^{*}}\sum_{x\in \mathbb{F}_{q}}\chi^{(t)}\biggl(\bigl(\mathrm{Tr}^{e}_{t}(x^{p^{\alpha}+1})-a\bigr)y\biggr)\\
      &+p^{-t-1}\sum_{z\in \mathbb{F}_{p}^{*}}\sum_{y\in \mathbb{F}_{p^{t}}^{*}}\sum_{x\in \mathbb{F}_{q}}\chi^{(t)}\biggl(\bigl(\mathrm{Tr}^{e}_{t}(x^{p^{\alpha}+1})-a\bigr)y\biggr)\chi^{(1)}\biggl(\bigl(\mathrm{Tr}^{e}_{1}(bx)-u\bigr)z\biggr)\\
      &=p^{e-t-1}+p^{-t-1}\biggl(M_{\alpha,b}(a,u)+L_{\alpha,b}(a,u)\biggr).
      \end{split}
      \end{equation*}
      We so far use $ \mathrm{Tr}^{e}_{1} $ to denote the absolute trace function $ \mathrm{Tr} $ in the proof. Note that $ \sum_{x\in \mathbb{F}_{q}}\sum_{z\in \mathbb{F}_{p}^{*}} \chi^{(1)}\biggl(\bigl(\mathrm{Tr}^{e}_{1}(bx)-u\bigr)z\biggr)=0 $. The actual lemma follows from Lemma \ref{lma8_ref}, Lemma \ref{lma9_ref} and (\ref{constant_kappa_general}).
   \end{proof}
   
   The following lemma gives the length of the codes $ \mathcal{C}_{D_{a}} $ of (\ref{linear_code_cda}), defined by (\ref{codelength_ref}):
       \begin{lemma10}\label{lma10_ref}
       Let $ a\in \mathbb{F}_{p^{t}} $. The cardinality of the defining set $ D_{a} $ of (\ref{definig_set}), i.e., the length of codewords of the linear codes $ \mathcal{C}_{D_{a}} $ of (\ref{linear_code_cda}), is determined by
       \begin{enumerate}[(I)]
       \item the first case: $ e/t \equiv 1\pmod 2 $.
       \begin{equation*}
       n_{\alpha}(a)=
       \begin{cases}
       p^{e-t}-1 &\quad \text{if }a=0\\
       p^{e-t}+\epsilon\kappa\eta^{(t)}(-a)p &\quad \text{if }a\ne 0.
       \end{cases}
       \end{equation*}
       \item The second case: $ e/t \equiv 0\pmod 2 $.
       \begin{equation*}
          n_{\alpha}(a)=
          \begin{cases}
          p^{e-t}+\epsilon\kappa_{2}(1-p^{-t})-1 &\quad \text{if }a=0\\
          p^{e-t}-\epsilon\kappa_{2}p^{-t} &\quad \text{if }a\ne 0.
          \end{cases}
       \end{equation*}
       \end{enumerate}
       \end{lemma10}
       \begin{proof}
       Let $ \delta:=-1 $ if $ a=0 $, and $ \delta:=0 $ if $ a\ne 0 $.
       Then, by the orthogonality of additive  character, (\ref{definig_set}), (\ref{codelength_ref}) and (\ref{MSum_ref}), we have
         \begin{equation*}
         \begin{split}
         n_{\alpha}(a)&=p^{-t}\sum_{x\in  \mathbb{F}_{q}}\biggl(\sum_{y\in  \mathbb{F}_{p^{t}}}\chi^{(t)}\biggl(y\bigl(\mathrm{Tr}_{t}^{e}(x^{p^{\alpha}+1})-a\bigr)\biggr)\biggr)+\delta\\
         &=p^{-t}\sum_{x\in  \mathbb{F}_{q}}\biggl(1+\sum_{y\in  \mathbb{F}_{p^{t}}^{*}}\chi^{(t)}\biggl(y\bigl(\mathrm{Tr}_{t}^{e}(x^{p^{\alpha}+1})-a\bigr)\biggr)\biggr)+\delta\\
         &=p^{e-t}+p^{-t}M_{\alpha}(a)+\delta.
         \end{split}
         \end{equation*}
         The actual lemma follows from Lemma \ref{lma8_ref} and (\ref{constant_kappa_general}).
       \end{proof}

  Recall that $ q=p^{e},d=\gcd(\alpha,e),1<t<d,t\mid d $. As $ e/d $ is odd,  $ f(x)=x^{p^{2\alpha}}+x\in \mathbb{F}_{q} $ is a permutation polynomial over $ \mathbb{F}_{q} $. For a given $ b\in \mathbb{F}_{q}^{*} $, let $ \gamma $ denote the unique solution of the equation $ f(x)=-b^{p^{\alpha}} $. Let $ Q_{t} $ denote all the squares of $ \mathbb{F}_{p^{t}}^{*} $, i.e., $ Q_{t}:=\lbrace x^{2}:x\in \mathbb{F}_{p^{t}}^{*}\rbrace $, and $ QN_{t}:=\mathbb{F}_{p^{t}}^{*}\setminus Q_{t} $. In order to determine the multiplicities of complete weight distributions of Theorem \ref{thm1_lab}-\ref{thm3_lab}, consider the following sets:
   \begin{enumerate}[(I)]
   \item The first case: $ e/t\equiv 1\pmod 2 $. \newline From (a) and (b), corresponding to $ e/t\equiv 1\pmod 2 $, of Lemma \ref{lma9_ref}, the multiplicities of complete weights of Theorem \ref{thm1_lab} are given by the cardinalities of the following sets:
   \begin{equation*}
      \begin{split}
      F^{(1)}_{1}&=\biggl\lbrace b\in \mathbb{F}_{q}^{*}:\mathrm{Tr}^{e}_{t}(\gamma^{p^{\alpha}+1})=0\biggr\rbrace,\\
      F^{(1)}_{2}&=\biggl\lbrace b\in \mathbb{F}_{q}^{*}:\mathrm{Tr}^{e}_{t}(\gamma^{p^{\alpha}+1})\in Q_{t}\biggr\rbrace,\\
      F^{(1)}_{3}&=\biggl\lbrace b\in \mathbb{F}_{q}^{*}:\mathrm{Tr}^{e}_{t}(\gamma^{p^{\alpha}+1})\in QN_{t}\biggr\rbrace.
      \end{split}
      \end{equation*}
      It is clear that in (\ref{cweformula_thm1}), $ F_{1}=\#F^{(1)}_{1}, F_{2}=\#F^{(1)}_{2}, F_{3}=\#F^{(1)}_{3}$.
      \newline From (c) and (d), corresponding to $ e/t\equiv 1\pmod 2 $, of Lemma \ref{lma9_ref}, the multiplicities of complete weights of Theorem \ref{thm2_lab} are given by the cardinalities of the following sets:
         \begin{equation*}
            \begin{split}
            F^{(2)}_{1}&=\biggl\lbrace b\in \mathbb{F}_{q}^{*}:\mathrm{Tr}^{e}_{t}(\gamma^{p^{\alpha}+1})=0\biggr\rbrace,\\
            F^{(2)}_{2}&=\biggl\lbrace b\in \mathbb{F}_{q}^{*}:\mathrm{Tr}^{e}_{t}(\gamma^{p^{\alpha}+1})\ne 0, \eta^{(t)}\bigl(\mathrm{Tr}^{e}_{t}(\gamma^{p^{\alpha}+1})\bigr)\ne \eta^{(t)}(a)\biggr\rbrace,\\
             F^{(2)}_{3}&=\biggl\lbrace b\in \mathbb{F}_{q}^{*}:\mathrm{Tr}^{e}_{t}(\gamma^{p^{\alpha}+1})\ne 0, \eta^{(t)}\bigl(\mathrm{Tr}^{e}_{t}(\gamma^{p^{\alpha}+1})\bigr)= \eta^{(t)}(a), \mathrm{Tr}^{t}_{1}\biggl(\sqrt{a\mathrm{Tr}^{e}_{t}(\gamma^{p^{\alpha}+1})}\biggr)=0\biggr\rbrace,\\
             F^{(2)}_{4}&=\biggl\lbrace b\in \mathbb{F}_{q}^{*}:\mathrm{Tr}^{e}_{t}(\gamma^{p^{\alpha}+1})\ne 0, \eta^{(t)}\bigl(\mathrm{Tr}^{e}_{t}(\gamma^{p^{\alpha}+1})\bigr)= \eta^{(t)}(a), \mathrm{Tr}^{t}_{1}\biggl(\sqrt{a\mathrm{Tr}^{e}_{t}(\gamma^{p^{\alpha}+1})}\biggr)\ne 0\biggr\rbrace.
            \end{split}
            \end{equation*}
    From (\ref{cweformula_thm3}), it is clear that $ F_{1}=\#F^{(2)}_{1}, F_{2}=\#F^{(2)}_{2}, F_{3}=\#F^{(2)}_{3}, F_{4}=\#F^{(2)}_{4}$.
   \item The second case: $ e/t\equiv 1\pmod 2 $. \newline From (a) and (b), corresponding to $ e/t\equiv 0\pmod 2 $, of Lemma \ref{lma9_ref}, the multiplicities of complete weights of Theorem \ref{thm3_lab} are given by the cardinalities of the following sets:
   \begin{equation*}
         \begin{split}
         F^{(3)}_{1}&=\biggl\lbrace b\in \mathbb{F}_{q}^{*}:\mathrm{Tr}^{e}_{t}(\gamma^{p^{\alpha}+1})=0\biggr\rbrace,\\
         F^{(3)}_{2}&=\biggl\lbrace b\in \mathbb{F}_{q}^{*}:\mathrm{Tr}^{e}_{t}(\gamma^{p^{\alpha}+1})\ne 0\biggr\rbrace.
         \end{split}
         \end{equation*}
   \end{enumerate}
   From (\ref{cweformula_thm5}), it is clear that $ F_{1}=  \#F^{(3)}_{1}, F_{2}=  \#F^{(3)}_{2}$.
   The following lemma gives the cardinalities of the above sets:
   \begin{lemma11}\label{lma11_ref}
    \[
     \begin{array}{ll}
     \#F^{(1)}_{1}=p^{e-t}-1,\\
     \#F^{(1)}_{2}=\frac{1}{2}(p^{t}-1)(p^{e-t}+\epsilon\kappa\eta^{(t)}(-1)p),  & \#F^{(1)}_{3}=\frac{1}{2}(p^{t}-1)(p^{e-t}-\epsilon\kappa\eta^{(t)}(-1)p),\\
     \#F^{(2)}_{1}=p^{e-t}-1, & \#F^{(2)}_{2}=\frac{1}{2}(p^{t}-1)(p^{e-t}-\epsilon\kappa\eta^{(t)}(-a)p),\\
     \#F^{(2)}_{3}=\frac{1}{2}(p^{t-1}-1)(p^{e-t}+\epsilon\kappa\eta^{(t)}(-a)p), & \#F^{(2)}_{4}=\frac{1}{2}(p-1)p^{t-1}(p^{e-t}+\epsilon\kappa\eta^{(t)}(-a)p),\\
     \#F^{(3)}_{1}=p^{e-t}+\epsilon\kappa p(p^{t}-1)-1,& \#F^{(3)}_{2}=(p^{t}-1)(p^{e-t}-\epsilon\kappa p).
     \end{array}
    \] 
   \end{lemma11}
   \begin{proof}
   We only prove $ \#F^{(2)}_{3} $ since the proofs for the remainder parts are similar.
   \begin{equation}\label{proof_lma11_ref}
   \begin{split}
  \#F^{(2)}_{3}&=\#\biggl\lbrace b\in \mathbb{F}_{q}^{*}:\mathrm{Tr}^{e}_{t}(\gamma^{p^{\alpha}+1})\ne 0, \eta^{(t)}\bigl(\mathrm{Tr}^{e}_{t}(\gamma^{p^{\alpha}+1})\bigr)= \eta^{(t)}(a), \mathrm{Tr}^{t}_{1}\biggl(\sqrt{a\mathrm{Tr}^{e}_{t}(\gamma^{p^{\alpha}+1})}\biggr)=0\biggr\rbrace\\
  &=\#\biggl\lbrace b\in \mathbb{F}_{q}^{*}:\mathrm{Tr}^{e}_{t}(\gamma^{p^{\alpha}+1})=k, k\in \mathbb{F}_{p^{t}}^{*}, \eta^{(t)}(k)= \eta^{(t)}(a), \mathrm{Tr}^{t}_{1}\bigl(\sqrt{ak}\bigr)=0\biggr\rbrace\\
  &=\#\biggl\lbrace x\in \mathbb{F}_{q}:\mathrm{Tr}^{e}_{t}(x^{p^{\alpha}+1})=k, k\in \mathbb{F}_{p^{t}}^{*}, \eta^{(t)}(k)= \eta^{(t)}(a), \mathrm{Tr}^{t}_{1}\bigl(\sqrt{ak}\bigr)=0\biggr\rbrace\\
  &=\sum_{m\in Q_{t}}\#\biggl\lbrace x\in \mathbb{F}_{q}:\mathrm{Tr}^{e}_{t}(x^{p^{\alpha}+1})=ma^{-1},  \mathrm{Tr}^{t}_{1}\bigl(\sqrt{m}\bigr)=0\biggr\rbrace\\
  &=\sum_{m\in Q_{t}}\sum_{x\in \mathbb{F}_{q} }\dfrac{1}{p^{t}}\biggl(\sum_{y\in \mathbb{F}_{p^{t}}}\chi^{(t)}\bigl((\mathrm{Tr}^{e}_{t}(x^{p^{\alpha}+1})-ma^{-1})y\bigr)\biggr)\dfrac{1}{p}\biggl(\sum_{z\in \mathbb{F}_{p}}\chi^{(1)}\bigl(\mathrm{Tr}^{t}_{1}\bigl(\sqrt{m}\bigr)z\bigr)\biggr)\\
  &=\dfrac{1}{p}\sum_{m\in Q_{t}}\sum_{z\in \mathbb{F}_{p}}\chi^{(1)}\bigl(\mathrm{Tr}^{t}_{1}\bigl(\sqrt{m}\bigr)z\bigr)n_{\alpha}(ma^{-1})
  =\dfrac{1}{2p}\sum_{y\in \mathbb{F}_{p^{t}}^{*}}\sum_{z\in \mathbb{F}_{p}}\chi^{(1)}\bigl(\mathrm{Tr}^{t}_{1}\bigl(\sqrt{y^{2}}\bigr)z\bigr)n_{\alpha}(y^{2}a^{-1})\\
  &=\dfrac{n_{\alpha}(a)}{2p}\sum_{y\in \mathbb{F}_{p^{t}}^{*}}\sum_{z\in \mathbb{F}_{p}}\chi^{(1)}\bigl(z\mathrm{Tr}^{t}_{1}(y)\bigr)
  =\dfrac{n_{\alpha}(a)}{2p}\sum_{y\in \mathbb{F}_{p^{t}}^{*}}\sum_{z\in \mathbb{F}_{p}}\chi^{(t)}\bigl(zy\bigr)\\
  &=\dfrac{n_{\alpha}(a)}{2p}\sum_{y\in \mathbb{F}_{p^{t}}^{*}}\biggl(1+\sum_{z\in \mathbb{F}_{p}^{*}}\chi^{(t)}\bigl(zy\bigr)\biggr)=\dfrac{n_{\alpha}(a)}{2p}\biggl((p^{t}-1)-(p-1)\biggr)
  =\dfrac{1}{2}(p^{t-1}-1)n_{\alpha}(a)\\
  &=\dfrac{1}{2}(p^{t-1}-1)\bigl(p^{e-t}+\epsilon\kappa\eta^{(t)}(-a)p\bigr). 
   \end{split}
   \end{equation}
   Notice that in the $ 5^{th} $ equation of (\ref{proof_lma11_ref}), according to (\ref{definig_set}), (\ref{codelength_ref}) and Lemma \ref{lma10_ref},
   \begin{equation*}
   \begin{split}
   n_{\alpha}(ma^{-1})&=\sum_{x\in \mathbb{F}_{q} }\dfrac{1}{p^{t}}\biggl(\sum_{y\in \mathbb{F}_{p^{t}}}\chi^{(t)}\bigl((\mathrm{Tr}^{e}_{t}(x^{p^{\alpha}+1})-ma^{-1})y\bigr)\biggr)\\
   &=p^{e-t}+\epsilon\kappa\eta^{(t)}(-ma^{-1})p=p^{e-t}+\epsilon\kappa\eta^{(t)}(-a)p=n_{\alpha}(a).
   \end{split}
   \end{equation*}
   \end{proof}
   
 Theorem \ref{thm1_lab}-\ref{thm3_lab} directly follow  from Lemma \ref{lma9_ref} and \ref{lma11_ref}.
   
  \section{Concluding remarks}
  In the present paper, two classes of linear codes with a few weights were constructed and their complete weight enumerators determined based on twisted Kloosterman sums. Let $ w_{min} $ and $ w_{max} $ denote the minimum and maximum nonzero weight of a linear code, respectively. It was shown (see \cite{BB17,BB40}) that any linear code can be used to construct secret sharing schemes with nice access structures provided $ w_{min}/w_{max}>(p-1)/p $.
  It can be verified that the linear codes of Theorem \ref{thm1_lab} and Theorem \ref{thm2_lab} fulfill this condition, so  can be used in the secret sharing.
  For the case that $ e/t\equiv 0\pmod 2 $ and $ a\in \mathbb{F}_{p^{t}}^{*} $, we were not able to determine the linear code $ \mathcal{C}_{D_{a}} $ due to the hard problem of explicitly evaluating the Kloosterman sum of (\ref{kloosterman_sum}). It should be interesting and challenging to try to solve the problem based on the works of  \cite{BB06,BB19} and the references therein, and then apply (\ref{kloosterman_sum}) in coding theory.

\end{document}